\documentclass[iop]{emulateapj}
\usepackage{graphicx}
\usepackage{nicefrac}
\usepackage{mathtools}
\usepackage{ulem}
\usepackage{color}
\usepackage[T1]{fontenc}
\usepackage{textcomp}

\shortauthors{Maga\~na, Motta, C\'ardenas, Verdugo \& Jullo}

\begin{document}
\title{A magnified glance into the Dark Sector: Probing cosmological models with strong lensing in A1689}
\shorttitle{Probing cosmological models with strong lensing in A1689}

\author{Juan Maga\~na$^{1}$}
\email{juan.magana@uv.cl}

\author{V. Motta$^{1}$}
\email{veronica.motta@uv.cl}

\author{V\'ictor H. C\'ardenas$^{1}$}
\email{victor.cardenas@uv.cl}

\author{T. Verdugo$^{1}$}
\email{tomasverdugo@gmail.com}

\author{Eric Jullo$^{2}$}
\email{eric.jullo@lam.fr}

\affil{$^{1}$Instituto de F\'{\i}sica y Astronom\'ia, Facultad
de Ciencias, Universidad de Valpara\'iso, Avda. Gran Breta\~na 1111, Valpara\'iso, Chile.}
\affil{$^{2}$Aix Marseille Universite, CNRS, LAM (Laboratoire d'Astrophysique de Marseille) UMR 7326, F-13388 Marseille, France}

\begin{abstract}
In this paper we constrain four alternative models to the
late cosmic acceleration in the Universe: Chevallier-Polarski-Linder
(CPL), interacting dark energy (IDE), Ricci holographic dark energy
(HDE), and modified polytropic Cardassian (MPC). Strong lensing (SL)
images of background galaxies produced by the galaxy cluster Abell
$1689$ are used to test these models. To perform this analysis we
modify the LENSTOOL lens modeling code. The value added by this
probe is compared with other complementary
probes: Type Ia supernovae (SNIa), baryon acoustic oscillations
(BAO), and cosmic microwave background (CMB). We found that the CPL
constraints obtained of the SL data are consistent
with those estimated using the other probes. The IDE constraints 
are consistent with the complementary bounds only if large errors 
in the SL measurements are considered. The Ricci HDE and MPC constraints are weak but 
they are similar to the BAO, SNIa and CMB estimations.
We also compute the figure-of-merit as a tool to
quantify the goodness of fit of the data. Our results suggest
that the SL method provides statistically significant constraints on
the CPL parameters but weak for those of the other models.
Finally, we show that the use of the SL measurements in galaxy
clusters is a promising and powerful technique to constrain
cosmological models. 
The advantage of this method is that
cosmological parameters are estimated by modelling the SL features
for each underlying cosmology. 
These estimations could be further improved by SL constraints coming from other galaxy clusters.
\end{abstract}


\section{Introduction}
The late cosmic acceleration, discovered by the Type IA supernovae
(SNIa) observations \citep{Perlmutter:1999, Riess:1998,
Schmidt:1998}, is the most intriguing feature of the Universe. What
gives origin this phenomenon is a big puzzle in modern cosmology.
There are two approaches that could drive the Universe to an
accelerated phase: an exotic component dubbed dark energy
\citep[DE,][]{cop06} and a modification of Einstein's gravity theory
\citep{modgrav}. In the DE scenario, the natural and most simple
model is the cosmological constant, $\Lambda$, associated to the
vacuum energy, and whose equation of state (EoS) parameter $w$, is
equal to $-1$. There are several cosmological observations beyond
SNIa data, such as the baryon acoustic oscillations (BAO) and
anisotropies of the cosmic microwave background (CMB) radiation,
supporting the cosmological constant as the nature of dark energy
\citep{Weinberg:2013}. Nevertheless, there are theoretical problems
associated to the cosmological constant: the fine-tuning problem,
i.e., its value is $\sim$120 orders of magnitude below the quantum
field theory prediction and the coincidence problem, that is, why
the DE density is similar to that of dark matter (DM) today
\citep{Weinberg:89}.

A straightforward way to solve these problems is by considering
models where the EoS evolves with time. Among the most studied
dynamical DE models are those involving scalar fields, for instance,
quintessence \citep{Wetterich:88, Peebles:88, Caldwell:1998,
quinta2}, phantom \citep{cal02,chi00}, quintom \citep{Guo:2005}, and
k-essence fields \citep{armendariza,armendarizb}. In addition, there
are many models in which the DE EoS is parameterized in terms of the
scale factor or redshift \citep{magana:2014}, for instance, the
well-known Chevallier-Polarski-Linder ansatz
\citep[CPL,][]{Chevallier:2000qy, Linder:2003nc}. The possibility of
interactions between the DM and DE are also considered by several
authors \citep{Bolotin:2015,CalderaCabral:2008bx, Valiviita:2009nu}.
These coupled models could alleviate both the coincidence problem
and the tension among different cosmological data
\citep{costa,He:2011,salvatelli,Valiviita:2015}. Other interesting
scenarios that have gained interest are the holographic dark energy
(HDE) models which are proposed in the context of a fundamental
principle of quantum gravity, so called the holographic principle
\citep{Cardenas:2010wx,Cardenas:2013moa,Hooft:93,susskind}. Although
some HDE models could alleviate the coincidence problem and are in
agreement with the cosmological data, they face many issues that
must be solved \citep{Cardenas:2013ela,delCampo14,zhang10}.

Thus, there are plenty of models with different theoretical
motivations, which are in agreement with some set of observational
data and explain the accelerated expansion in the Universe
\citep{Li:2013}. To discriminate among all these scenarios it is
common to put constraints on their parameters using the distance
modulus from SNIa, the CMB anisotropies, and BAO
\citep{Nesseris:2005,Nesseris:2006er,Shi:2012}. Many of the current
data analysis are performed assuming a fiducial $\Lambda$ cold DM
model. Therefore, to improve the cosmological parameter estimation
and to avoid biased constraints due to the assumption of a model, it
is necessary to acquire high-precision data and to develop new
complementary cosmological techniques, such as cosmography, which studies
a set of observables of the Universe's kinematics \citep[see][and references therein]{Gruber:2014}.

Several authors have shown that the strong gravitational effect can
be used as a powerful probe to test cosmological models
\citep[e.g.,][]{cao, Chen:2013,collet,Cardenas:2013moa,Jullo:2010,Lubini:2014}.
Strong lensing (SL) occurs whenever the light rays of a source are
strongly deflected by the lens, producing multiples images of the
background source. The position of these images depend on the
properties of the lens mass distribution. As the Einstein radii also
depends on the cosmological model, the SL observations have been
used to derive constraints on the DM density parameter,
$\Omega_{DM}$, and the EoS for alternative DE models \citep[see for
example][]{biesiada_a,biesiada_b,Biesiada:2011zza}. In these
previous works, the alternative cosmological models are tested by
comparing (for the lens systems) the theoretical ratio of the angular
diameter distances with an observable. This observable is typically
estimated assuming a particular lens model along the standard
cosmological paradigm. Nevertheless, the best way should be test the
cosmological model by reconstructing the lens model with that
underlying new cosmology. A pioneer work using a parametric reconstruction was
performed by \citet{Jullo:2010} to probe a flat constant $w$CDM
model using the SL measurements in the Abell$1689$ (A$1689$) galaxy
cluster. They found that the DE EoS estimated with this technique is
in agreement with those obtained using CMB and BAO. 
Recently \citet{Lubini:2014} investigated a novel non-parametric
SL lens modelling to determine cosmological parameters. They applied 
this procedure using synthetic lenses and showed that it
is possible to infer unbiased constraints from the assumed cosmological parameters.
Therefore, SL modeling in galaxy clusters is a powerful and complementary method
to put constraints on cosmological parameters \citep[see also][and
references therein]{DAloisio:2011}.

In this paper we extend the previous analysis of \citet{Jullo:2010}
to test four alternative models using the SL measurements of A$1689$
galaxy cluster. We investigate whether this technique is able to put
narrow constraints on the dark parameters and the consistency of
them with those provided by the SNIa, BAO and CMB data. The paper is
organized as follows: in the next section \S \ref{sec:data} we
briefly describe the data used to constrain the cosmological
parameters. In section \S \ref{sec:models} we introduce the
framework for a flat Universe and the cosmological models to be
tested. In section \S \ref{sec:method} we define the method to
obtain the constrains for each data set We present the results in
section \S \ref{sec:results} and discuss them in section \S
\ref{sec:discussion}. Finally, we give our conclusions in section \S
\ref{sec:conclusions}.

\section{The data} \label{sec:data}
The following four data sets are used to test the alternative
cosmological models: SL measurements in A\,$1689$ galaxy cluster,
SNIa, BAO, and CMB.

\textit{SL in A$1689$.-} A$1689$ is among the richest clusters given
the number density of galaxies in its core, one of the most luminous
of galaxy clusters in X-ray wavelengths \citep{Ebeling1996}, it
displays an incredible large number of arc systems
\citep[see][]{Limousin2007}, and it has been studied using
gravitational lensing by several authors \citep[e.g.,][and
references therein]{Limousin2007,Limousin2013,Diego2015,Umetsu2015}.
A$\,1689$ was previously used by \citet{Jullo:2010} to
simultaneously constrain the cluster mass distribution and DE EoS
employing a SL parametric model. We refer the interested reader to
that paper, for a detailed description of the methodology to select
the final catalog of multiple-image systems used to perform their
analysis. In our present work, we are using the same catalog, which
consist on 28 images derived from $12$ families with spectroscopic
redshift range $1.15<z_{S}<4.86$.

\textit{SNIa.-} We use the sample presented by
\citet{Ganeshalingam:2013mia} consisting on $586$ SNIa in the
redshift range $0.01-1.4$ which considers $91$ points of the Lick
Observatory Supernova Search (LOSS) sample
\citep{Ganeshalingam:2010}.

\textit{BAOs.-} Baryon acoustic oscillation signature is a useful
standard ruler to constrain the expansion of the Universe by the
distance-redshift measurements from clustering of galaxies with
large scale surveys \citep{Blake:2003,Seo:2003}. The BAO
measurements considered in our analysis are obtained from the
Six-degree-Field Galaxy Survey (6dFGS) BAO data
\citep{Beutler2011:6dF}, the WiggleZ experiment
\citep{Blake2011:wigglez}, the Sloan Digital Sky Survey (SDSS) Data
Release 7 (DR7) BAO distance measurements \citep{Percival2010:dr7},
the Baryon Oscillation Spectroscopic Survey (BOSS) SDSS Data Release
9 BAO distance measurements \citep[SDSS DR9,][]{Anderson2012:dr9},
and the most recent BAO distance estimations from Data release $11$
(DR11) of the BOSS of SDSS \citep{Delubac2015}.

\textit{CMB.-} The CMB power spectra is sensitive to the distance to
the decoupling epoch, at redshift $z_{*}$, via the locations of
peaks and the acoustic oscillations. The CMB measures two distance
ratios related to the decoupling epoch: the acoustic scale
$l_{A}(z_{*})$, and the shift parameter $R(z_{*})$. A quick way to
confront a cosmological model with the CMB data without run a
Bayesian global analysis of the power spectra is via the fitting of
both distances \citep{wang:2006,Wright:2007}. The cosmological
constraints estimated using this method are consistent with those
obtained of the full analysis \citep{Komatsu:2009}. Moreover,
although these distance posterior are computed assuming an
underlying cosmology, \citet{Li:2008} has demonstrated that these
quantities are almost independent on the input DE models. We include
CMB information by using the $l_{A}$, $R$, and $z_{*}$ posteriors
from the WMAP 9-yr measurements \citep{Hinshaw:2013}.

\section{Cosmological models} \label{sec:models}

A flat Friedmann-Lema\^itre-Robertson-Walker (FLRW) Universe with
scale factor $a$ and Hubble parameter $H(a)=\dot{a}/a$ is
considered. For each cosmological model we use the following
components: a source of cosmic acceleration, cold DM and a radiation
fluid ($\gamma$, photons and relativistic neutrinos). For this
universe, the comoving distance from the observer to redshift $z$ is
given by
\begin{equation}
r(z)=\frac{c}{H_0}\int_0^z \frac{dz'}{E(z')}.
\label{eq:rz}
\end{equation}
\noindent
where $E(z)=H(z)/H_{0}$ and $H_{0}=H(0)$.
The angular diameter distance for a source at redshift $z$ is
\begin{equation}
D_{A} (z) =  \frac{r(z)}{1+z}.
\label{eq:dA}
\end{equation}
\noindent 
Since we are interested in the ability of SL measurements to constraints the parameters
related to the cosmic acceleration and DE, during our analysis we
set the Hubble parameter $H_{0}=70\,\mathrm{km\,s^{-1}Mpc^{-1}}$ and
$\Omega_{DM0}=0.27$. 
The current density parameter for radiation is
$\Omega_{\gamma 0}=2.469\times10^{-5}h^{-2}(1+0.2271 N_{eff})$,
where $h=H_{0}/100\, \mathrm{kms}^{-1}\mathrm{Mpc}^{-1}$ and the
number of relativistic species is set to $N_{eff}=3.04$
\citep{Komatsu:2011}. At low redshifts,
$\Omega_{\gamma}\,(\sim10^{-5})\ll\Omega_{DM},\Omega_{DE}\,(\sim1)$,
thus we neglect this term when we use the A$1689$ SL measurements,
but it is taken into account on the other data sets.
The assumption of these fiducial values 
allows direct model comparisons because each model
only has two free parameters related to DE.
\citet{Bayliss:2015} produced magnification maps for the Hubble Frontier Fields (HFF) galaxy clusters 
\footnote{http://www.stsci.edu/hst/campaigns/frontier-fields/} using priors in
$\Omega_{DM0}$ and $H_{0}$. They obtain that varying the input cosmological parameters
results in significant differences in the magnification maps. 
Nevertheless, the influence of the $H_{0}$ and its uncertainty in the lens reconstruction
is subdominant because it cancels out when calculating the distance ratio (see Eqs. \ref{eq:rz}, \ref{eq:dA}
and \ref{eq:Xi}). On the other hand, the parameter estimation in the SL modelling
could be slightly biased due to different choices of $\Omega_{DM0}$.
However, this bias is not statistically significative.
Thus, for simplicity, we did not take into account
the cosmological parameter ($\Omega_{DM0}$, $H_{0}$) uncertainties in the SL lens modelling.

We choose the following alternative cosmological models:
Chevallier-Polarski-Linder, interacting dark energy (IDE), Ricci HDE, 
and modified polytropic Cardassian (MPC). In the following
subsections we present the chosen models and the reasons for
selecting them.

\subsection{Chevallier-Polarski-Linder model}
A natural extension to the $\Lambda$CDM scenario which could solve the coincidence
problem is to allow the DE EoS to vary with time or redshift via some parametrization.
One of the most popular function is the CPL parametrization
\citep{Chevallier:2000qy, Linder:2003nc} given by

\begin{equation}
w(z) = w_{0} + w_{1} \frac{z}{1+z},
\label{eq:wcpl}
\end{equation}
where $w_{0}=w(0)$, $w_{1}=w'(0)$ are constants to be fitted by the
data. The $E(z)$ function for a FLRW Universe where the DE EoS is
parametrized and expressed as

\begin{eqnarray}
E^{2}(z)&=&\Omega_{\gamma0}(1+z)^{4}+\Omega_{DM0}(1+z)^{3} + \nonumber\\
&&\left( 1 - \Omega_{DM0}- \Omega_{\gamma0}\right)f(z),
\end{eqnarray}
where
\begin{equation}
f(z)=\mathrm{exp}\left(3\int^{z}_{0}\frac{1+w(z)}{1+z}\mathrm{dz}\right).
\label{eq:fz}
\end{equation}
The substitution of the Eq. (\ref{eq:wcpl}) in (\ref{eq:fz}) results in:
\begin{equation}
f(z)=(1+z)^{3(1 + w_{0} + w_{1})} \exp \left[ - \frac{3w_{1}z}{1+z}\right].
\end{equation}
Therefore, $E^{2}(z,\Theta)$ for the CPL parametrization reads as
\begin{eqnarray}
&&E^{2}(z,\Theta)= \Omega_{\gamma0}(1+z)^{4} + \Omega_{DM0}(1+z)^{3} +\nonumber\\
&&\left(1- \Omega_{DM0}-\Omega_{\gamma0}\right) (1+z)^{3(1 + w_{0} + w_{1})} \exp \left[ - \frac{3w_{1}z}{1+z}\right],\qquad
\label{eq:Ecpl}
\end{eqnarray}
\noindent where $\Theta=(w_{0},w_{1})$ is the vector of the free
parameters to be fitted by the data. The CPL parametrization is the
fiducial model proposed by the Dark Energy Task Force (DETF) to
study the cosmic acceleration \citep{Albrecht:2006}. Therefore, the
Eq. (\ref{eq:Ecpl}) has been widely used to put constraints on
$w_{0}$ and $w{1}$ \citep[see for example][]{Su:2011,Shi:2012}.

\subsection{Interacting Dark Energy model}

In IDE models there is a relation between the DE energy density
($\rho_{DE}$), and the DM energy density ($\rho_{DM}$) that could
alleviate the cosmic coincidence problem. The general approach
introduce a $Q$ strength term in the right-side of the continuity
equations for the dark components as follows
\citep{Amendola:1999er,CalderaCabral:2008bx,Cai:2004dk,Dalal:2001dt,Guo:2007zk,
Valiviita:2009nu}:
\begin{eqnarray}
\dot{\rho_{DM}} &+& 3H\rho_{DM} =Q,\nonumber\\
\dot{\rho_{DE}} &+& 3H\left(1 + w_{x}\right)\rho_{DE}= -Q,
\label{ec:ide}
\end{eqnarray}
\noindent where $w_{x}$ is the EoS of IDE. There are many choices
for the phenomenological energy exchange term $Q$. One of them is to
assume $Q$ to be proportional to the Hubble rate, $H$, times either
the energy densities or their sum or some other combination of the
energy densities. We consider $Q=\delta H \rho_{DM}$, being $\delta$
a constant to be fitted by the data \citep[it is equivalent to
$Q=3\delta H \rho_{DM}$ studied by][]{Cao:2011cg,costa,He:2011}. A
positive $\delta$ describes an energy transfer or a decay of DM to
DE and a negative $\delta$ corresponds to an energy transfer from DE
to DM. The $E^{2}(z)$ function \citep[see its calculation
in][]{Bolotin:2015,Guo:2007zk} for this IDE reads as
\begin{eqnarray}
&&E^{2}(z,\Theta) = \Omega_{\gamma0}(1+z)^{4}+ \nonumber\\
&&(1 -\Omega_{DM0}-\Omega_{\gamma0})(1+z)^{3(1 + w_{x})} + \nonumber\\
&&\frac{\Omega_{DM0}}{\delta + 3 w_{x}} \left[ \delta(1+z)^{3(1 + w_{x})} +  3 w_{x}(1+z)^{3 - \delta} \right],
\label{eq:EzIDE}
\end{eqnarray}
\noindent where the free parameters to be constrained by the data
are $\Theta=(w_{x},\delta)$. The Eq. (\ref{eq:EzIDE}) has been
considered in flat \citep{Cao:2011cg,costa,Guo:2007zk,He:2011} and
non-flat \citep{Shi:2012} models to put constraints on $w_{x}$ and
$\delta$ using several cosmological data.

\subsection{Holographic dark energy with Ricci scale and CPL parametrization}

Many dark energy models invoke the holographic principle (HP) which
states that the number of degrees of freedom of a physical system
should be finite and it should scale with its bounding area rather
than with its volume \citep{Hooft:93, Fischler:1998, susskind}. In
HDE it is required that the total energy in a region of size $L$
should not exceed the mass of a black hole of the same size, thus
the HDE energy density satisfies $L^{3} \rho_{HDE}\leq M_{p}^{2} L$
\citep{Cohen:1999}. This expression imposes a relationship between
the ultraviolet (UV, related to the vacuum energy) and infrared (IR,
related to large-scale of the Universe) cutoffs. By saturating this
inequality, we obtain the following DE energy density
\begin{equation}
\rho_{HDE}=\frac{3c_{H}^{2}M_{p}^{2}}{L^{2}},
\end{equation}
\noindent
where the numerical constant $c_{H}$ is related with the degree of saturation of the previous inequality.
Therefore, the DE energy becomes dynamical and the fine-tunning and coincidence problems
could be solved. There are several ways to choose the IR cut-off, for example,
the Hubble horizon, or the event horizon \citep{delCampo11}. Here, we consider $L^{2}=6/\mathcal{R}$,
where $\mathcal{R}$ is the Ricci scalar defined as $\mathcal{R}=6(2H^{2}+\dot{H})$ \citep{Gao:2009,delCampo13}.
We also consider that the DM and DE interact with each other obeying Eqs. (\ref{ec:ide}).
Following the work by \citet{delCampo11}, we parametrize the EoS with the CPL ansatz (\ref{eq:wcpl}).
The $E(z,\Theta)$ parameter (see Appendix \ref{sec:appHDE}) for this model is the following
\begin{eqnarray}
E^{2}(z,\Theta)&=&\frac{\left(1+z\right)^{4}}{1+x_{0}+y_{0}}\left(\frac{1+x_{0}}{f}\right)^{2\alpha}\times\nonumber\\
&&\left[f+y_{0}\left(\frac{f}{1+x_{0}}\right)^{2\alpha}\right],
\label{eq:Ehde}
\end{eqnarray}
where $x_{0}=\nicefrac{\Omega_{DM0}}{\Omega_{HDE0}}$, $y_{0}=\nicefrac{\Omega_{\gamma0}}{\Omega_{HDE0}}$,
and $\Omega_{HDE0}=1-\Omega_{DM0}-\Omega_{\gamma0}$. The function $f$ and the exponent $\alpha$ are:
\begin{equation}
f=1+x_{0}+ z(1+3w_{1}+x_{0}),
\end{equation}

\begin{equation}
\alpha=1-\frac{3(w_{0}+w_{1})}{2(1+3w_{1}+x_{0})}.
\end{equation}
\noindent
The free parameter vector to be fitted by the data is $\Theta=(w_{0},w_{1})$.
A similar model without the radiation component was tested by \citet{Cardenas:2013moa}.
We present a new analytical solution for the Ricci HDE model with CPL parametrization
which includes a relativistic fluid.

\subsection{Modified Polytropic Cardassian model}

The original Cardassian model was introduced by \citet{Freese:2002}
to explain the accelerated expansion of the universe without DE.
Motivated by braneworld theory, this model modifies the Friedmann
equation as $H^{2}=8\pi G \rho_{m}/3 + B\rho_{m}^{n}$, where $\rho_{m}$ is the total matter density.
The second term in the right hand side, known as the Cardassian term,
drives the universe to an accelerated phase if the exponent $n$ satisfies $n<2/3$.
\citet{gondolo} introduced a simple generalization of the Cardassian
model, the modified polytropic Cardassian, by introducing an
additional exponent $q$ \citep[see also][]{Wang:2003}. The modified
Friedmann equation with this generalization can be written as
\begin{equation}
H^{2} = \frac{8\pi G}{3} \rho_{m} \left[ 1 + \left( \frac{\rho_{Card}}{\rho_{m}} \right)^{q(1-n)} \right]^{1/q},
\label{eq:Hmpc}
\end{equation}
where $\rho_{Card}$ is the characteristic energy density with $n<2/3$ and $q>0$. At early
times, the universe is expanded according to the canonical Friedmann
equation. However, at late times, the Cardassian term dominates
driving the universe to an accelerated expansion phase.
The equation (\ref{eq:Hmpc}) reduces to the $\Lambda$CDM model for $q = 1$ and $n= 0$.
Introducing a radiation term, the dimensionless $E^{2}(z,\Theta)$ parameter reads as:
\begin{eqnarray}
&&E^{2}(z,\Theta)=\Omega_{r}(1+z)^{4} + \Omega_{m}(1+z)^{3} \times \nonumber\\
&&\left[ 1 + \left( \left( \frac{1-\Omega_{r}}{\Omega_{m}}\right)^{q}  - 1 \right)(1+z)^{3q(n - 1)}  \right]^{1/q},
\label{eq:EzMPC}
\end{eqnarray}
where the free parameter vector to be fitted by the data is $\Theta=(q,n)$.
The flat MPC model (Eq. \ref{eq:EzMPC}) has been studied by several authors using different data without
the radiation component \citep{feng_card} and also with a curvature term \citep{Shi:2012}.

\section{The method} \label{sec:method}

In this section we explain how the cosmological parameters are
estimated for each different observational data set, and we also
define the merit functions for each one of them.

\subsection{Strong lensing}

In the SL regime, the light beams are deflected so strongly that they can result in the observation
of several distorted images of a background source. The positions of the multiple images depend
significantly on the characteristics of the lens mass distribution. Since the image positions are also
related to the angular diameter distance ratios between the lens, source and observer, they retain information
about the underlying cosmology. In particular, this dependence of the lensing models on the geometry
can be used to derive constraints on the DM density parameter and the DE EoS \citep[see][]{Jullo:2010}.

The cosmological models discussed in the section \S \ref{sec:models} were implemented in
LENSTOOL\footnote{\tt This software is publicly available at:
http://projets.lam.fr/projects/lenstool/wiki} ray-tracing code,
which uses a Bayesian  Markov chain Monte Carlo (MCMC) method
\citep{Jullo:2007}. The model fitting is carry out taking into account
the cosmological sensitivity of the angular size-redshift relation,
when sources are at distinct redshifts \citep{Link1998}. Using this method,
the angular diameter distance ratios for $2$ images from different
sources defines the 'family ratio' \citep[see][for a detailed discussion]{Jullo:2010},
for which the constraints on cosmological parameters could be obtained:

\begin{equation}
\Xi(z_1,z_{s1},z_{s2},\Theta) = \frac{D(z_1,z_{s1})}{D(0,z_{s1})}\frac{D(0,z_{s2})}{D(z_1,z_{s2})},
\label{eq:Xi}
\end{equation}
\noindent where $\Theta$ is the vector of cosmological parameters to be fitted,
$z_1$ is the lens redshift, $z_{s1}$ and $z_{s2}$
are the two source redshifts, and $D(z_1,z_2)$ is the angular
diameter distance, calculated through Eq. (\ref{eq:rz}) and Eq.
(\ref{eq:dA}).

We computed the models performing the optimization in the source
plane. We solved the lens equation in the source plane because it is 
computationally more efficient and we checked with some models that source and image plane results were similar.
Note that differences can appear for complex clusters with irregular shape 
\citep[e.g. MACSJ0717.5+3745,][]{Limousin:2012}, but this is not the case with Abell 1689.
Every lensing mass model (regardless of the DE model)  has a
total of $21$ free parameters, and consists of two large-scale
potentials, a galaxy-scale potential for the central brightest
cluster galaxy (BCG), and includes the modeling of $58$ of the
brightest cluster galaxies.

For each of the image systems ($12$ families, see \S \ref{sec:data})
with n images, we determine the goodness of fit for a particular set
of model parameters defining a $\chi^{2}$:

\begin{equation}
\chi^2 = \sum_{i=1}^n \frac{\left[M(\vec{\beta_i} - \langle \vec{\beta} \rangle) \right]^2}{\sigma_i^{2}},
\end{equation}

\noindent where $\beta_i$ is te source plane position corresponding
to image $i$, $\langle \vec{\beta} \rangle$ is the family
barycenter, $M$ is the magnification tensor, and $\sigma_i$ is the
total error \citep[see][]{Jullo:2007}. The total $\chi^2$ is computed summing over the whole
set of families.

\subsection{Type Ia Supernovae}
The SNIa samples give the distance modulus as a function of redshift
$\mu_{obs}(z)$ and its error $\sigma_{\mu}$. Theoretically, the
distance modulus is computed as
\begin{equation}
\mu(z)=5\log_{10}[d_L(z)/\texttt{Mpc}]+\mu_{0},
\label{eq:mu}
\end{equation}
where $\mu_{0}$ is a nuisance parameter which depends on
the absolute magnitude of a fiducial SN Ia and the Hubble parameter.
The $\mu(z)$ is a function of the cosmological model through the luminosity distance
(measured in Mpc)
\begin{equation}\label{dlzf}
d_L(z)=(1+z)r(z),
\end{equation}
where $r(z)$ is given by Eq. (\ref{eq:rz}). By marginalizing over $\mu_{0}$,
we obtain $\chi^2_{SNIa}=A-B^{2}/C$, where
\begin{eqnarray}
A&=&\sum_{i=1}^{586}\frac{[\mu(z_i)-\mu_{obs}(z_i)]^2}{\sigma_{\mu_{i}}^2},\nonumber\\
B&=&\sum_{i=1}^{586}\frac{\mu(z_i)-\mu_{obs}(z_i)}{\sigma_{\mu_{i}}^2},\nonumber\\
C&=&\sum_{i=1}^{586}\frac{1}{\sigma_{\mu_{i}}^2}.
\label{eq:chisnia}
\end{eqnarray}
The SNIa constraints can be estimated by minimizing the $\chi^2_{SNIa}$.

\subsection{BAOs measurements}

The 6dFGS BAO estimated the distance ratio $d_{z}=0.336\pm0.015$ at
$z=0.106$ \citep{Beutler2011:6dF}, where
\begin{equation}
d_{z} = \frac{r_s(z_d)}{D_V(z)}.
\label{eq:dz}
\end{equation}
The comoving sound horizon, $r_s(z)$, is defined as
\begin{equation}
r_s(z) = c \int_z^\infty \frac{c_s(z')}{H(z')}dz',
\end{equation}
where the sound speed is $c_s(z) = 1/\sqrt{3(1+\bar{R_b}/(1+z)}$, with
$\bar{R_b} = 31500\, \Omega_{b}h^2(T_{CMB}/2.7\,\mbox{K})^{-4}$, $\Omega_{b}$
is baryonic density parameter, and $T_{CMB}$
is the CMB temperature \citep[$2.726$K for WMAP 9-yr,][]{Hinshaw:2013}.
The distance scale $D_V$ is defined as
\begin{equation}
D_V(z)=\frac{1}{H_0}\left[(1+z)^2D_A(z)^2\frac{cz}{E(z)}\right]^{1/3},
\end{equation}
where $D_A(z)$ is the angular diameter distance given by Eq. (\ref{eq:dA}).

The redshift $z_d$ at the baryon drag epoch is fitted with the
formula proposed by \citet{Eisenstein98},
\begin{equation}
z_d =
\frac{1291(\Omega_{m}h^2)^{0.251}}{1+0.659\,(\Omega_{m}h^2)^{0.828}}[1+b_1(\Omega_b
h^2)^{b_2}],
\end{equation}
where
\begin{eqnarray}
b_1 &=& 0.313\left(\Omega_{m}\,h^2\right)^{-0.419}\left[1+0.607\left(\Omega_{m}\,h^2\right)^{0.674}\right], \\
b_2 &=& 0.238\left(\Omega_{m}\,h^2\right)^{0.223}.
\end{eqnarray}

Therefore, the $\chi^2$ for the BAO data point from 6dFGS is
\begin{equation}
\chi^2_{\scriptscriptstyle 6dFGS} =
\left(\frac{d_z-0.336}{0.015}\right)^2.
\end{equation}

The WiggleZ BAO estimated three points for the acoustic parameter $A(z)$ \citep{Eisenstein05}
\begin{equation}
A(z) = \frac{D_V(z)\sqrt{\Omega_{m}H_0^2}}{cz}.
\label{eq:Az}
\end{equation}
The observational data are $\bar{A}_{obs} = (0.474,0.442,0.424)$ for the effective redshifts $z=0.44,0.6$ and 0.73 respectively.

Thus, the $\chi^2$ for the WiggleZ BAO data is given by
\begin{equation}
\chi^2_{\scriptscriptstyle WiggleZ} = (\bar{A}_{obs}-\bar{A}_{th})
C_{\scriptscriptstyle WiggleZ}^{-1} (\bar{A}_{obs}-\bar{A}_{th})^T,
\end{equation}
where $\bar{A}_{th}$ denotes the theoretical value for the acoustic parameter $A(z)$
and $\bar{A}_{obs}$ the observed one.
The inverse covariance
$C_{\scriptscriptstyle WiggleZ}^{-1}$ is given by
\begin{equation}
C_{\scriptscriptstyle WiggleZ}^{-1} = \left(
\begin{array}{ccc}
1040.3 & -807.5 & 336.8\\
-807.5 & 3720.3 & -1551.9\\
336.8 & -1551.9 & 2914.9
\end{array}\right).
\end{equation}

Similarly, for the SDSS DR7 BAO distance measurements,
the $\chi^2$ can be expressed as
\begin{equation}
\chi^2_{\scriptscriptstyle DR 7} =
(\bar{d_{z}}_{obs}-\bar{d_{z}}_{th})C_{\scriptscriptstyle DR7}^{-1}(\bar{d_{z}}_{obs}-\bar{d_{z}}_{th})^T,
\label{ap:xsdss}
\end{equation}
where $\bar{d_{z}}_{obs} = (0.190195,0.1097)$ are the data at $z=0.2$, and $0.35$ respectively \citep{Percival2010:dr7}.
Here $\bar{d_{z}}_{th}$ denotes the theoretical distance ratio given by Eq. (\ref{eq:dz}).
The inverse covariance matrix $C_{\scriptscriptstyle DR7}^{-1}$ reads as
\begin{equation}
C_{\scriptscriptstyle DR7}^{-1} = \left(
\begin{array}{cc}
30124 & -17227\\
-17227 & 86977
\end{array}\right).
\end{equation}

The SDSS DR9 estimated the distance ratio $d_{z}=0.0732\pm0.0012$ at $z=0.57$ \citep{Anderson2012:dr9}.
For this BAO data point, the $\chi^2$ function is given by
\begin{equation}
\chi^2_{\scriptscriptstyle DR9} =
\left(\frac{d_z-0.0732}{0.0012}\right)^2.
\end{equation}

The most recent measured position of the BAO peak from SDSS DR11 determine
$D_{H}/r_{d}=9.18\pm0.28$ at $z=2.34$, where $D_{H}=c/H$ and $r_{d}=r_{s}(z_{d})$ \citep{Delubac2015}.
We compute the $\chi^2$ for this point as
\begin{equation}
\chi^2_{\scriptscriptstyle DR11} =
\left(\frac{D_H/r_{d}-9.18}{0.28}\right)^2.
\end{equation}

Therefore, the total $\chi^{2}$ function for the BAO measurements is
\begin{equation}
\chi^2_{BAO} = \chi^2_{\scriptscriptstyle6dFGS}+\chi^2_{\scriptscriptstyle WiggleZ} +
\chi^2_{\scriptscriptstyle DR7} + \chi^2_{\scriptscriptstyle DR9} + \chi^2_{\scriptscriptstyle DR11}.
\label{eq:chibao}
\end{equation}
The BAO constraints can be estimated by minimizing the Eq. (\ref{eq:chibao}).

\subsection{CMB}
We use the following WMAP 9-yr distance posterior \citep{Hinshaw:2013} for a flat $\Lambda$CDM Universe:
$l_{A^{obs}}=302.40$, $R^{obs}=1.7246$, $z_{*}^{obs}=1090.88$, and the inverse covariance matrix

\begin{equation}
C_{WMAP9}^{-1} = \left(
\begin{array}{ccc}
3.182 & 18.253 & -1.429\\
18.253 & 11887.879 & -193.808\\
-1.429 & -193.808 & 4.556
\end{array}\right).
\label{eq:Cwmap9}
\end{equation}
The acoustic scale is defined as
\begin{equation}
l_A = \frac{\pi r(z_*)}{r_s(z_*)},
\label{eq:lA}
\end{equation}
and the redshift of decoupling $z_*$ is given by \citep{Hu:1996},
\begin{equation}
z_* = 1048[1+0.00124(\Omega_b h^2)^{-0.738}]
[1+g_1(\Omega_{m}h^2)^{g_2}],
\end{equation}
\begin{equation}
g_1 = \frac{0.0783(\Omega_b h^2)^{-0.238}}{1+39.5(\Omega_b
h^2)^{0.763}},
 g_2 = \frac{0.560}{1+21.1(\Omega_b h^2)^{1.81}}.
\end{equation}
The shift parameter is defined as \citep{Bond:1997}
\begin{equation}
R = \frac{\sqrt{\Omega_{m}H_{0}^2}}{c} r(z_{*}).
\end{equation}

Thus, the CMB constraints can be estimated by minimizing
\begin{equation}\label{cmbchi}
 \chi^2_{CMB} = X^TC_{CMB}^{-1}X,
\end{equation}
where
\begin{equation}
 X =\left(
 \begin{array}{c}
 l_A^{th} - l_A^{obs} \\
 R^{th} -  R^{obs}\\
 z_*^{th} - z_{*}^{obs}
\end{array}\right),
\end{equation}
and the superscripts $th$ and $obs$ refer to the theoretical and observational values respectively.

\section{Results} \label{sec:results}

The parameters of the alternative models are determined by minimizing the $\chi^{2}$
function for each data set. For all models, first we calculate the minimum values using the
SL data with the priors described by \citet{Jullo:2010}.
Then, we estimate the constraints using the  SNIa, BAO, and CMB, comparing the different data sets.
Furthermore we could use a refined $\chi^{2}_{min}$
criteria, for instance, the Akaike information criterion (AIC), and the Bayesian information criterion (BIC),
to discern which model is prefered by the data. However, since all the tested models have only
two free parameters, the information provided by the $\chi^{2}_{min}$ values is sufficient and
AIC and BIC criteria does not provide further information.\\

\begin{figure}
\includegraphics[width=0.5\textwidth]{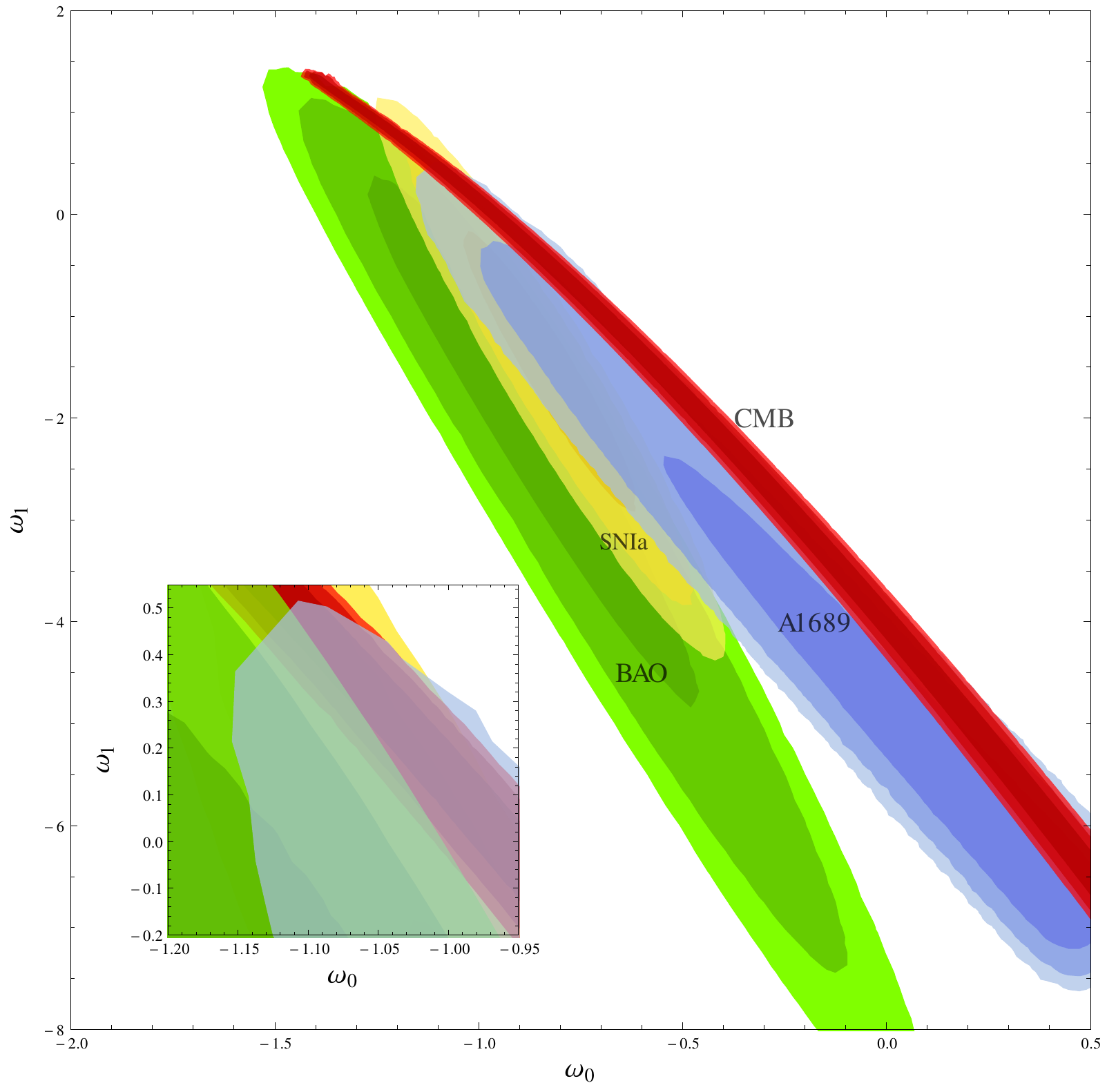}
\caption{The $1\sigma$, $2\sigma$ and $3\sigma$ contours of the CPL model parameters $w_{0}$ and $w_{1}$
obtained from different data sets: Abell\,1689 (blue), SNIa (yellow), CMB (red) and BAO (green).
The inset shows the region where the contours overlap}
\label{fig:cpl}
\end{figure}

\textit{CPL .-} The best fits on the EoS parameters $w_{0}$ and $w_{1}$ for the CPL model
and the estimated $\chi^{2}$ using each data set are listed in Table \ref{tab:cpl}.
Note that the limits derived for the A$1689$ SL data are in tension with those obtained
with the SNIa, BAO, and 9yr-WMAP data. Actually, the A$1689$ constraint on $w_{0}$
is positive implying no cosmic acceleration.  The Fig. \ref{fig:cpl} shows the marginalized contours at
$1\sigma$, $2\sigma$ and $3\sigma$ for the CPL parameters. The inset
shows the region where the different contours overlap.

\begin{deluxetable}{lllll}
\tablecolumns{5}
\tablecaption{CPL model\label{tab:cpl}}
\tablehead{
\colhead{Data set} & \colhead{$\chi^{2}_{min}$} & \colhead{FoM}\tablenotemark{a}&\colhead{$w_{0}$} & \colhead{$w_{1}$}}
\startdata
A$1689$&  $264.9$& $8.20$ & $0.43\pm0.48$&  $-6.45^{+3.60}_{-0.36}$ \\
SNIa&  $574.13$&  $24.41$ &$-0.82\pm0.14$ &$-1.51\pm0.91$ \\
BAO & $3.77$& $7.89$  &$-0.94\pm0.26$&  $-1.55\pm1.72$ \\
CMB&  $0.363$& $21.54$ & $-0.59\pm0.58$&  $-1.38\pm2.36$
\enddata
\tablecomments{Best fits for the $w_{0}$ and $w_{1}$ CPL parameters estimated from the
SL measurements in A$1689$, SNIa, BAO and CMB.}
\tablenotetext{a}{We define the FoM in section \ref{subsec:meritsl}.}
\end{deluxetable}

\textit{IDE .-} We have carried out two different analysis using the SL data:
the first one, A$1689$M1, with the (astrometric) errors in the image positions of the systems
previously used by \citet{Jullo:2010} (also used in the other three models
of this paper) and the other, A$1689$M2, in which we have set the errors
as five times the values of the fiducial model, i.e., five times the errors in model A$1689$M1,
so that we obtain a reduced $\chi^2 \simeq 1$. These large errors take into account 
other possible sources of uncertainties in the SL measurements such as systematic errors 
due to the complexities in the mass distribution and the line-of-sight structures 
\citep[e.g.][]{DAloisio:2011,Jullo:2010}.
In \S \ref{subsec:IDE_discussion} we will resume the discussion again.

\begin{figure}
\centering
\includegraphics[width=0.5\textwidth]{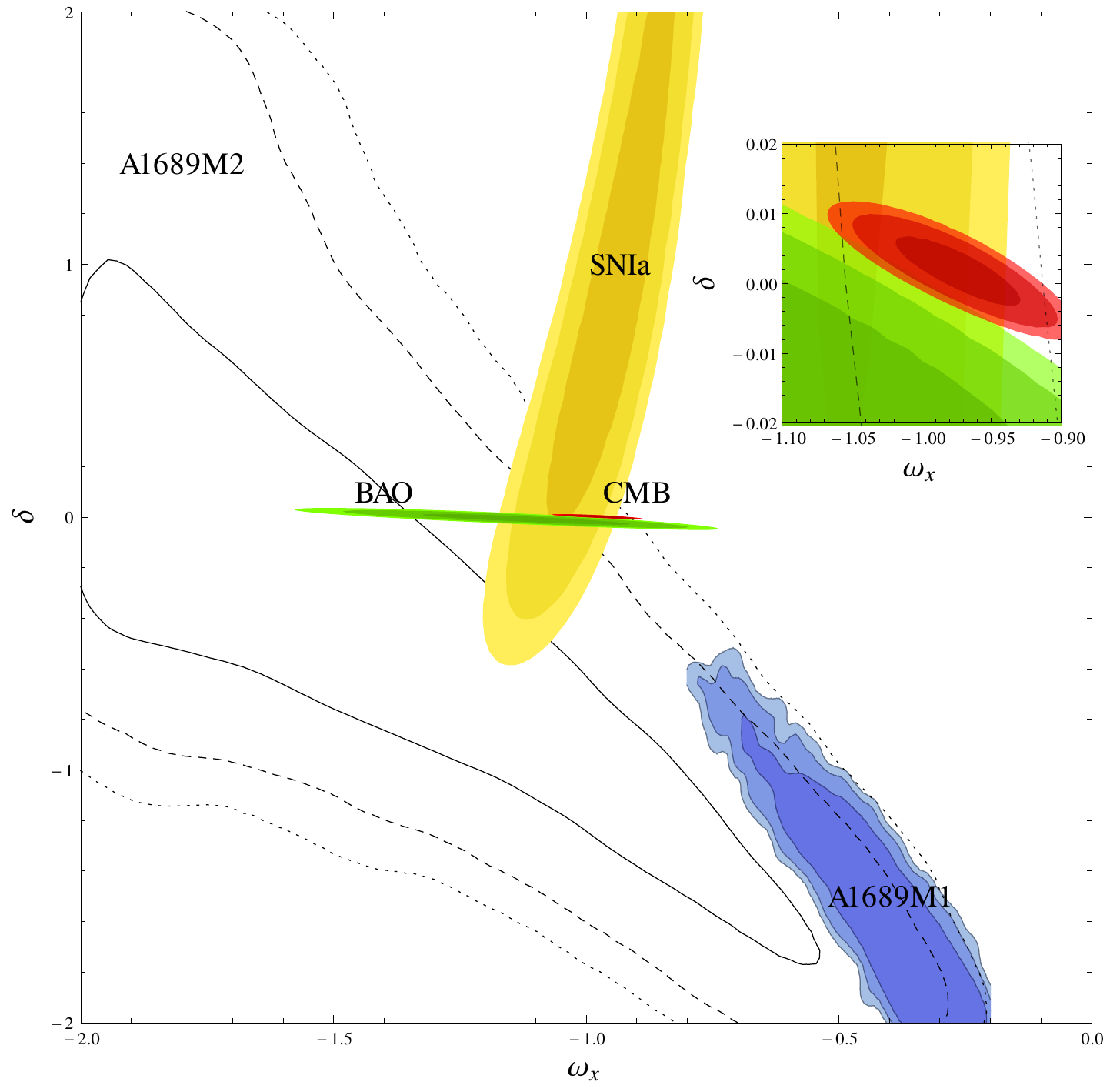}
\caption{The $1\sigma$, $2\sigma$ and $3\sigma$ contours of the IDE model parameters $\delta$ and $w_{x}$
obtained from different data sets: A$1689$M1 (blue), SNIa (yellow), CMB (red) and BAO (green).
The contours for the run A$1689$M2 considering large errors are: $1\sigma$ in solid line,
$2\sigma$, dashed line, and $3\sigma$ dotted line. The inset shows the region where the
of SNIa, BAO and CMB contours overlap with those of the A$1689$M2 analysis.}
\label{fig:ide}
\end{figure}

The best fits on the $w_{x}$ and $\delta$ parameters for both runs of the IDE model
and the estimated $\chi^{2}$ using each data set are listed in Table \ref{tab:ide}.
Note that A$1689$M1 constraints on $w_{x}$ and $\delta$ are in disagreement with the estimations
of the other cosmological tests. In the second analysis, considering larger errors on the SL data,
we obtain that the best fit on $w_{x}$ is in agreement at $1\sigma$ with the others data.
The Fig. \ref{fig:ide} shows the marginalized $1\sigma$, $2\sigma$ and $3\sigma$
confidence contours in the plane $w_{x} - \delta$ for each data set.
The inset shows the region where the A$1689$M2 contours overlap with SNIa, BAO, and CMB contours.

\begin{deluxetable}{lllll}
\tablecolumns{5}
\tablecaption{IDE model\label{tab:ide}}
\tablehead{
\colhead{Data set} & \colhead{$\chi^{2}_{min}$} & \colhead{FoM} & \colhead{$w_{x}$} & \colhead{$\delta$}}
\startdata
A$1689\mbox{M}1$ & $256.7$ & $127.063$ & $-0.32\pm0.07$ & $-2.0^{+0.30}_{-0.0}$ \\
A$1689\mbox{M}2$ & $25.9$ & $4.55$ & $-1.53\pm0.42$ & $-0.21\pm0.80$ \\
SNIa & $574.95$ & $38.76$ &$-0.95\pm0.08$& $0.77\pm0.69$ \\
BAO & $4.61$ & $1060.52$ &$-1.10\pm0.13$ & $-0.0093\pm0.014$ \\
CMB & $0.081$ & $18488.1$ &$-0.97\pm0.02$ & $-0.0017\pm0.003$
\enddata
\tablecomments{Best fits for the $w_{x}$ and $\delta$ IDE parameters estimated from the
SL measurements in A$1689$, SNIa, BAO and CMB.}
\end{deluxetable}

\begin{figure}
\centering
\includegraphics[width=0.5\textwidth]{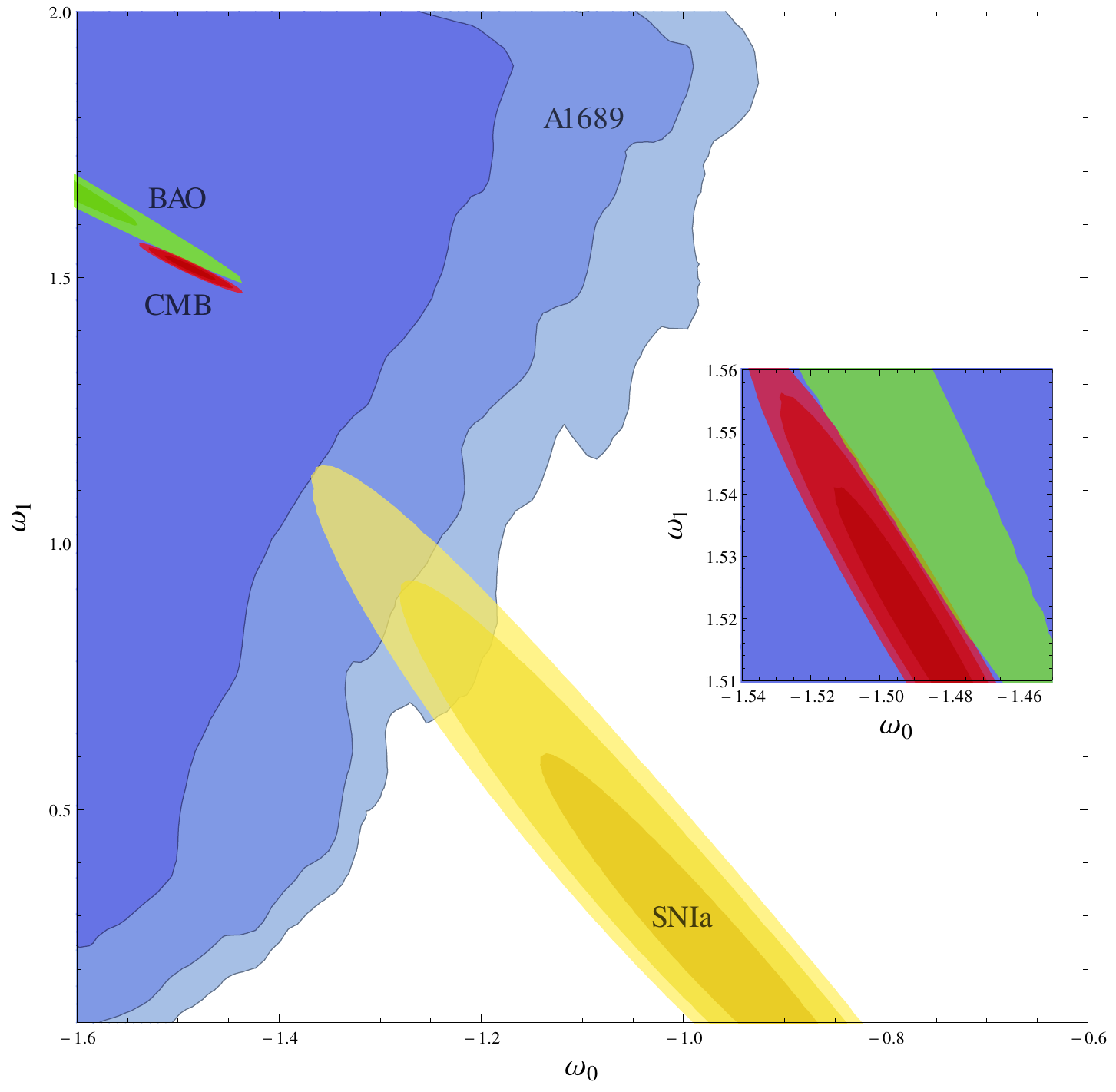}
\caption{The $1\sigma$, $2\sigma$ and $3\sigma$ contours of the Ricci HDE model parameters $w_{0}$ and $w_{1}$
obtained from different data sets: Abell1689 (blue), SNIa (yellow), CMB (red) and BAO (green).
The inset shows the region where the contours overlap.}
\label{fig:holo}
\end{figure}

\textit{Ricci HDE .-} The The best fits on $w_{0}$ and $w_{1}$ parameters and
the $\chi^{2}$ obtained using each data set are shown in Table \ref{tab:holo}.
Note that A$1689$ best fits are consistent with those of the CMB data. Nevertheless,
there is a tension between these values and the BAO and SNIa constraints.
Figure \ref{fig:holo} shows the marginalized confidence contours at $1\sigma$, $2\sigma$, and
$3\sigma$ in the parameters space $w_{0}$-$w_{1}$. The inset shows the region where the contours overlap.

\begin{deluxetable}{lllll}
\tablecolumns{5}
\tablecaption{HDE model\label{tab:holo}}
\tablehead{
\colhead{Data set} & \colhead{$\chi^{2}_{min}$} & \colhead{FoM} &\colhead{$w_{0}$} & \colhead{$w_{1}$}}
\startdata
A$1689$ & $279.82$ & $24.85$ &$-1.60^{+0.13}_{-0.0}$ & $1.97^{+0.01}_{-0.66}$ \\
SNIa & $575.135$ & $153.89$ &$-0.96\pm0.10$& $0.21\pm0.22$ \\
BAO & $5.79$ & $241.52$ &$-2.03\pm0.21$ & $2.10\pm0.21$ \\
CMB & $0.081$ & $14725.7$ &$-1.48\pm0.01$ & $1.51\pm0.01$
\enddata
\tablecomments{Best fits for the $w_{0}$ and $w_{1}$ Ricci HDE parameters estimated from the
SL measurements in A$1689$, SNIa, BAO and CMB.}
\end{deluxetable}

\begin{figure}
\centering
\includegraphics[width=0.5\textwidth]{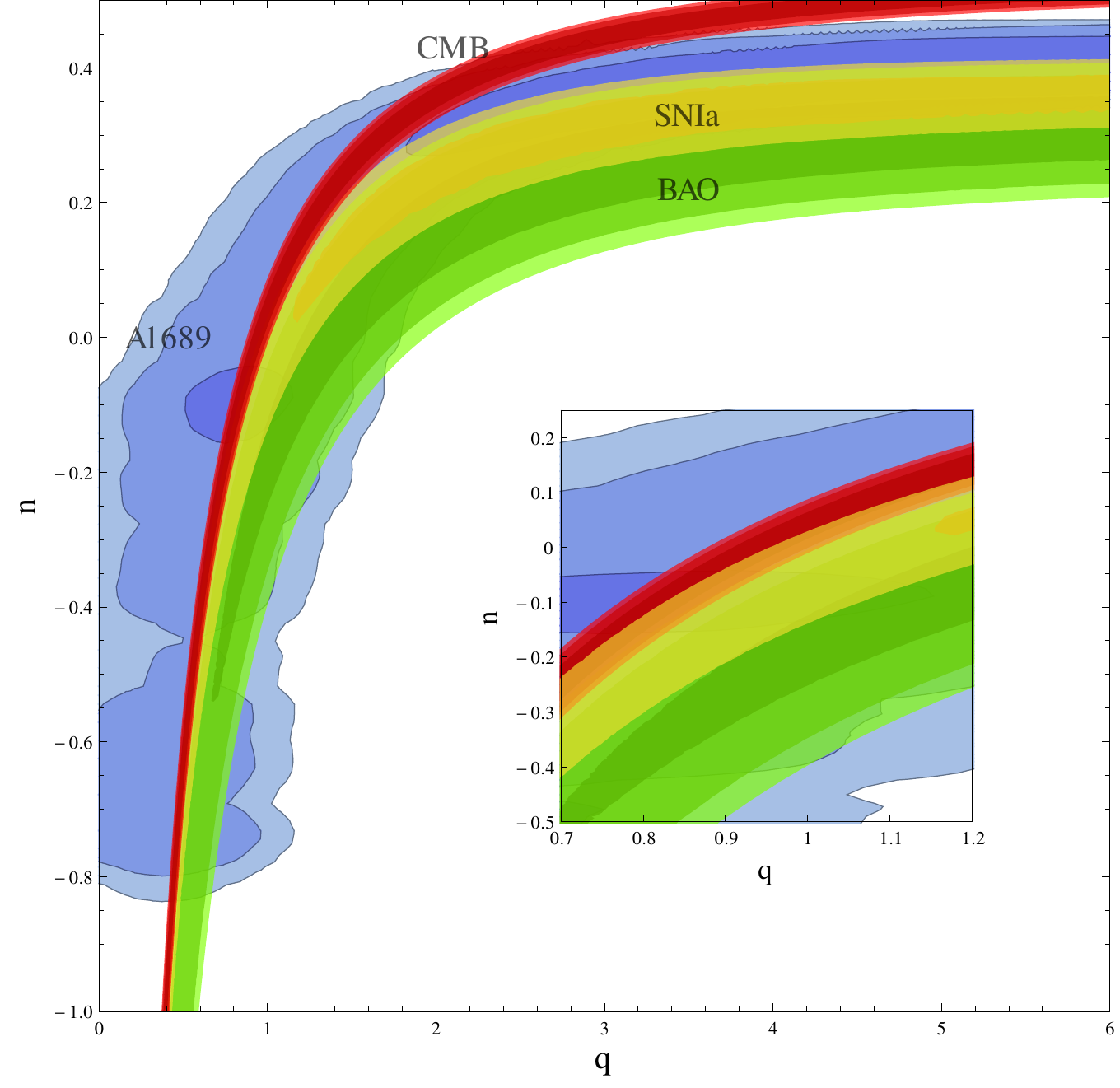}
\caption{The $1\sigma$, $2\sigma$ and $3\sigma$ contours of the MPC model parameters $q$ and $n$
obtained from different data sets: A$1689$ (blue), SNIa (yellow), CMB (red) and BAO (green).
The inset shows the region where the contours overlap.}
\label{fig:card}
\end{figure}

\textit{MPC .-} The best fits on the $q$ and $n$ parameters
and the estimated $\chi^{2}$ using each data set are listed in Table \ref{tab:car}.
Note that the A$1689$ constraints on $q$ and $n$ are consistent with the limits given by CMB data.
Although, these SL and CMB best fits are in tension with the estimations obtained with SNIa and BAO observations,
the errors for the $q$ parameter are statistically larger, thus the constraints for each data are consistent at $1\sigma$.
Figure \ref{fig:card} shows the marginalized confidence contours at $1\sigma$, $2\sigma$, and
$3\sigma$ in the parameters space $q$-$n$. An interesting aspect of the SL contours
is that the A$1689$ data provided two $1\sigma$ regions.
The inset shows a region where the different contours overlap.
\begin{deluxetable}{lllll}
\tablecolumns{5}
\tablecaption{MPC model\label{tab:car}}
\tablehead{
\colhead{Data set} & \colhead{$\chi^{2}_{min}$} & \colhead{FoM} &\colhead{$q$} & \colhead{$n$}}
\startdata
A$1689$ & $266.7$ & $2.54$& $5.2\pm2.25$ & $0.41\pm0.25$ \\
SNIa & $574.52$ & $18.69$ &$3.20\pm2.19$& $0.32\pm0.08$ \\
BAO & $3.59$ & $7.97$ &$3.29\pm3.30$ & $0.26\pm0.13$ \\
CMB & $0.363$ & $37.63$ &$4.52\pm3.27$ & $0.49\pm0.05$
\enddata
\tablecomments{Best fits for the $q$ and $n$ Cardassian parameters estimated from the
SL measurements in A$1689$, SNIa, BAO and CMB.}
\end{deluxetable}

\section{Discussion} \label{sec:discussion}

\subsection{CPL}

By combining WMAP-9yr data, the measurements of CMB from Atacama Cosmology
Telescope (ACT) and the South Pole Telescope (SPT), BAO points, and $H_{0}$ measurements, \citet{Hinshaw:2013}
estimated $w_{0} = -1.17^{+0.13}_{-0.12}$, $w_{1} = 0.35^{+0.50}_{-0.49}$.
These values are in concordance at $1\sigma$ within our limits although there is a
significative tension in the constraints on $w_{1}$. Furthermore, the SL estimations are consistent with
$w_{0} = -1.04^{+0.72}_{-0.69}$, $w_{1} < 1.32$ and the approximated range $-1.2<w_{0}<-0.5$, $-1.8<w_{1}<0.2$
obtained by \citet{PlanckXVI, PlanckXIV:2015} respectively. 
Note that the SL contours (see Fig. \ref{fig:cpl}) are analogous to those obtained with the other data.
The overlapped region suggests the cutoffs $w_{0}\approx-1.05$ and $w_{1}\sim0.2$, which
is consistent with the cosmological constant. It is worthy to mention that the CPL contours,
using the different data sets, are similar to those computed by \citet{PlanckXIV:2015}.

\subsection{IDE} \label{subsec:IDE_discussion}
For the SL, BAO, and CMB data the $\delta$ constraints
are negative suggesting an energy transfer from DM to DE opposite to the SNIa constraint. Moreover,
there is a tension with the contraints provided by the BAO and CMB data which favor no evidence of dark interactions.
The Fig. \ref{fig:ide} shows that the contours obtained from the run A$1689$M1
do not overlap with the plots computed using the other data.
Nevertheless, the contours derived of the analysis A$1689$M2
are consistent and orthogonal with those of SNIa, BAO and CMB data.
These contours, computed assuming the aforementioned large errors in the image position of the SL systems,
indicates that it is necessary to take into account all the sources of errors (including systematic errors) 
in the SL models to avoid severe biases in the calculations of DE constraints \citep{DAloisio:2011}.
Furthermore, several authors have shown that the line-of-sight structure is a significant source 
of uncertainty in the SL mass modelling, and consequently, in the observed image positions 
\citep{Bayliss:2014, Host:2012, Jaroszynski:2014, McCully:2014, Zitrin:2015}. 
While our assumed errors could be slightly overestimated \citep[see e.g.][]{Zitrin:2015},
\citet{DAloisio:2011} showed that the observational errors (in the case of space-based imaging) 
are typically an order of magnitude lower than potential modelling errors.
Thus, a realistic SL model should take these uncertainties into account
by using large errors in the position of multiple images.

Although the data from A$1689$M2 give weak constraints on the IDE parameters,
they provide significant evidence of interactions between the DM and DE. The overlapped region
of the Fig. \ref{fig:ide} suggests the cutoffs $w_{x}\approx-1$ and $-0.005<\delta<0.005$, which
is consistent with the cosmological constant and no interactions in the dark sector.
Similar constraints on $\delta$ was also obtained by \citet{He:2011}\footnote{Notice that in the \citet{costa,He:2011,Cao:2011cg} works,
they define a coupling constant $\delta_{m}$, which is equivalent to $\delta/3$}
using the WMAP seven-year data and others cosmological observations
\citep[see also the consistency with the bounds estimated by][]{Cao:2011cg}.
In a recent paper by \citet{costa}, the authors put constraints on this IDE model
using Planck data in combination with SNIa, BAO, and Hubble parameter measurements.
They found $w_{x}\approx-1.65$ for CMB data alone and $w_{x}\approx-1.25$ from the joint analysis.
In addition, they found slightly evidence of energy transfer from DE to DM ($\delta\sim0.006$).

\subsection{Ricci HDE}
Recently, \citet{Cardenas:2013moa} tested this model without the radiation fluid
performing a joint analysis of SL, BAO, SNIa, and $H(z)$ data \citep[see also][]{delCampo13}.
They found $w_{0}=-1.27^{+0.12}_{-0.13}$ and $w_{1}=0.99^{+0.30}_{-0.26}$.
These values are consistent at $2\sigma$ with our constraints obtained of A$1689$ SL measurements.
The Ricci HDE confidence contours (see Figure \ref{fig:holo}) show that SL data produce weaker constraints
on $w_{0}$ and $w_{1}$. Moreover, the $1\sigma$ SL contour overlap with the CMB bounds
and only at $2\sigma$, and $3\sigma$ for the BAO and SNIa data.
It is worthy to note that the SNIa contours are only overlap with those derived of the SL probe.
Several authors have pointed out that this tension between the SNIa constraints together
with those of BAO and CMB tests could be due to the choice of priors on the DM density parameter,
statistical and systematic errors in the data sets, the choice of different SNIa light-curve fitters, etc
\citep[see for example][]{Escamilla-Rivera:2012, Gong:2010, Lazkoz:2008, Li:2011, magana:2014,
Nesseris:2005,Perivolaropoulos:2009}.
The tension of the Ricci HDE constraints derived from
several data sets will be further investigated in a forthcoming work.

\subsection{MPC}
Several authors have tested the MPC model using different data sets.
For instance, the Cosmic All-Sky Survey (CLASS) lensing sample have
been used by \citet{Alcaniz:2005} to obtain the constraints $q=0.05$ and $n=-2.32$
which are in tension with our SL fits. \citet{wang:2009} estimated, using
SNIa, BAO, CMB, Hubble parameter measurements and the gas mass fraction in galaxy clusters,
$q=0.824^{+0.750}_{-0.622}$, and $n=-0.091^{+0.331}_{-1.908}$
which are in agreement with our confidence contours (see Figure \ref{fig:card}).
In addition, our limits are similar at $1\sigma$ to
$q=0.480^{+2.020}_{-0.080}$, $n=-0.600^{+0.980}_{-0.450}$ computed by \citet{feng_card}
using the combination of SNIa (Constitution sample), BAO, and 5-yrs WMAP data.
By combining SNIa, BAO, CMB and gamma-ray burst data, \citet{Liang:2011} found
$q=0.76^{+0.36}_{-0.58}$ and $n=-0.16^{+0.25}_{-3.26}$ in accordance with one of the $1\sigma$ SL contour
as well as those of CMB, and SNIa. Recently, \citet{Li:2012} use different SNIa samples together with
BAO and CMB data to put the constraints $q=1.098^{+1.015}_{-0.465}$ and $n=0.014^{+0.364}_{-0.964}$
which are consistent with our limits at $1\sigma$. 
Note that the SL contours are similar in shape and orientation to those obtained with the other data.
In addition, our confidence contours are similar to those computed by \citet{Liang:2011,Li:2012,wang:2009}
The overlapped region (see inset of Figure \ref{fig:card}) suggests the cutoffs $0.45<q<1.05$ and $-0.8<n<0.05$.
The weak constraints on the MPC parameters obtained with the different data sets 
do not provide strong evidence of modifications to the Friedmann
equations, hence of cosmic acceleration without DE.

\subsection{Merit of the SL method} \label{subsec:meritsl}

As we showed and discussed above, the SL technique provides complementary
constraints to the standard cosmological probes. It is important
to stress that the determination of which cosmological model
is favored by the data, mainly by the SL measurements, is far from the scope of the present work
(the current data do not allow us to undertake such detailed analysis). Nevertheless,
by considering standard errors in the SL data, the IDE model gives the lowest value of the SL $\chi^{2}_{min}$,
therefore it is the favored by the A$1668$ SL measurements. However, as discussed in
\S \ref{subsec:IDE_discussion}, the SL constraints for this IDE model are in disagreement
with the those of SNIa, BAO, and CMB. The CPL model is the second one
preferred by the SL data. 

\begin{figure}
\includegraphics[width=0.5\textwidth]{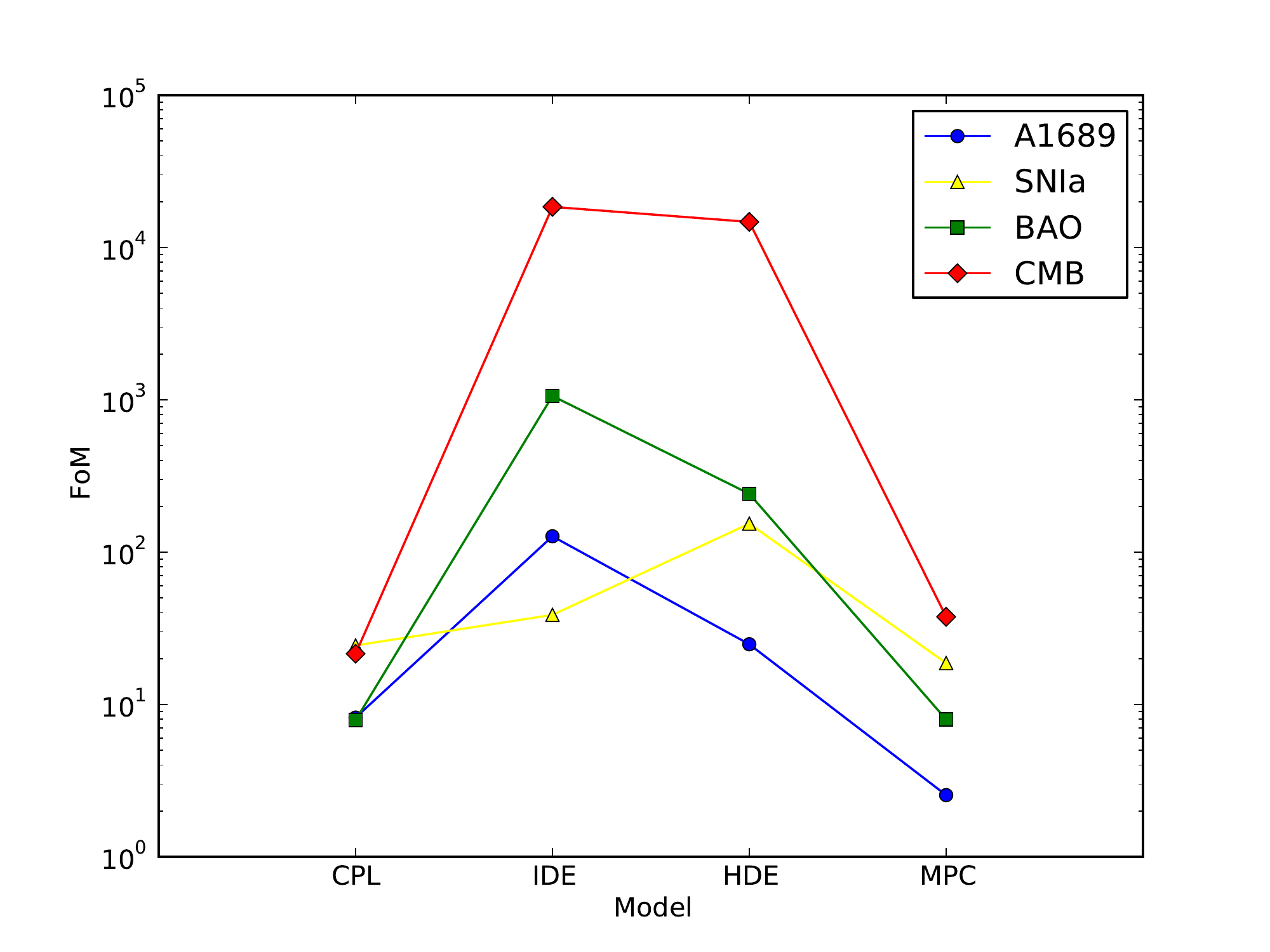}
\caption{FoM for each model using A$1689$, SNIa, BAO and CMB data.\\}
\label{fig:fom}
\end{figure}

Another useful tool to quantify the ability of each observational data 
to constrain the cosmological parameters is considering the Figure-of-Merit
\citep[FoM, ][]{Albrecht:2006,Su:2011,wang:2008}.
The DETF defined the FoM for the CPL model as the inverse of the area enclosed
by the 95\% confidence level contour of $(w0, wa)$ \citep{Albrecht:2006}.
\citet{wang:2008} introduced a more general definition given by
\begin{equation}
\mbox{FoM}=\frac{1}{\sqrt{\det \mbox{Cov} (f1,f2,f3,...)}},
\end{equation}
\noindent
where $\mbox{Cov} (f1,f2,f3,...)$ is the covariance matrix of the
cosmological parameters {$f_{i}$}.
Larger FoM means stronger constraints on the parameters
since it corresponds to a smaller error ellipse.
We have computed the FoM of the cosmological models for each data set.
The results are shown in the third column of the Tables \ref{tab:cpl}-\ref{tab:car}.
For a more intuitive comparision, we show in Figure \ref{fig:fom}
the values of the FoM for each model using each data set.
Note that for the CPL model, the SL probe gives slightly more stringent constraints
than the BAO test. In the case of A$1689$M1, although the SL FoM is $\sim3.3$ times greater than
that SNIa, the constraints obtained from this lens model are inconsistent with the others tests.
For the A$1689$M2 analysis with large errors on the SL measurements, the SL FoM has
the lowest value, indicating that this technique provides weak constraints on the IDE parameters.
The FoMs for the HDE and MPC models from the A$1689$ data are also the lowest
when compared with the other cosmological probes. Although it is very difficult to compare
the SL FoM for different models, our computations suggest that
the SL technique provides statistically significant constraints on the CPL parameters but
a weak for the other models. 

In spite of the general low FoM for the SL probe compared with the other data sets, it is important to remark
that we only use the measurements of one galaxy cluster, namely, A$1689$. To improve
the SL constraints on the parameters of the alternative cosmological models
we need to consider the SL measurements in other galaxy clusters.
This work is a first test of the capability of this technique to constrain unusual cosmological models.
\citet{DAloisio:2011} showed that using the data from 10 simulated galaxy clusters,
each one with 20 multiply imaged families, the estimated constraints on the parameters of
the $w$CDM and CPL models are improved.
We plan to extend this analysis using the coming data from future surveys such as
The Frontier Fields (FF) program of the Hubble space telescope. 
In addition, it is crucial that future efforts also take
into account for additional uncertainties in the lens modelling due to line-of-sight 
structure and other systematic errors \citep{Bayliss:2014, Host:2012, Jaroszynski:2014, McCully:2014}. 
Finally, it is important to point out that the cosmological constraints obtained in this work
could be affected by other unknown systematics \citep{Bayliss:2014, Bayliss:2015, Zitrin:2015}
such as the SL modelling technique.

\section{Conclusions} \label{sec:conclusions}
One of the main goals of observational cosmology is to
elucidate what gives origin to the late cosmic acceleration in the Universe.
A wide set of theoretical models have been proposed to explain this cosmic feature \citep{Li:2013}
and need to be tested with observational data \citep{Albrecht:2006,Lazkoz:2008, Nesseris:2005, Nesseris:2006er}.
In this paper, we put constraints on four alternative cosmological models:
Chevallier-Polarski-Linder parametrization, interacting dark energy, Ricci
holographic dark energy and modified polytropic Cardassian.
We mainly focus on a powerful and not fully exploited technique
which uses the strong lensing measurements in A$\,1689$ galaxy cluster \citep{Jullo:2010}.
The advantage of the method presented here is that the cosmological parameters are estimated
by modelling the SL features for each underlying cosmology.
Additionally, we use the SNIa, BAO and CMB signal as complementary probes.
We have shown that for the CPL model the SL method provide constraints
in agreement with those estimated with the other probes. We performed
two analysis for the IDE model, one with standard errors in the SL measurements and the other
with larger errors to takes into account other sources of uncertainties 
\citep{Bayliss:2014, DAloisio:2011, Host:2012, Jaroszynski:2014, McCully:2014, Zitrin:2015}.
We found that the limits on the IDE parameters derived of the standad error analysis are in
disagreement with the standard tests. Moreover, the confidence contours
do not overlap with those of SNIa, BAO, and CMB. Nevertheless, if larger errors in the SL measurements
are considered, the SL estimations are consistent with the constraints obtained from other probes.
Therefore, underestimating the total error can lead to erroneous constraints on the parameters of the IDE model. 
For the Ricci HDE, the SL data give weak constraints on the DE EoS parameters.
In addition, we also found a tension between the bounds obtained from SNIa, BAO and CMB data.
Finally, the estimations for the MPC parameters using the SL test are similar to the  SNIa, BAO, and CMB
contraints. We calculate also the figure-of-merits to quantify the goodness
of fitting using the different data. We found that in general the SL constraints
are weak when compared with other tests. Also, contours not always overlap with each other, 
suggesting some systematic errors in the  models of the observables that remain to be investigated.
Nevertheless, it is worthy to note that we use only data from one galaxy cluster. The cosmological constraints
could be improved if more SL data are used \citep{DAloisio:2011}.
Our results show that this is a powerful technique that will be used in the future, when more data
are available, in particular those for the HFF clusters.

\acknowledgments
We thank the anonymous referee for thoughtful suggestions.
J.~M. acknowledges the support from ESO Comit\'e Mixto, Gemini $32130024$, ECOS-CONICYT C$12$U$02$, and
the hospitality of LAM where part of this work was done.
V.~M. acknowledges support from FONDECYT $1120741$, ECOS-CONICYT C12U02,
and Centro de Astrof\'{\i}sica de Valpara\'{\i}so.
V.~C. acknowledges support from FONDECYT Grant $1110230$ and DIUV $13/2009$.
T.~V. thanks Dr. V.~Motta for the kind invitation to work in Valpara\'{\i}so,
as well as the staff of the Instituto de F\'{\i}sica y Astronom\'{\i}a of the Universidad de Valpara\'{\i}so.
E.J. acknowledges the support of CNES. This work was granted access to the HPC resources 
of Aix-Marseille Universit\'e financed by the project Equip@Meso (ANR-10-EQPX-29-01) 
of the program \guillemotleft Investissements d'Avenir\guillemotright\, supervised by the Agence Nationale pour la Recherche. 

\appendix
\section{The $E(z)$ analytical solution for the Ricci HDE model with a
radiation component} \label{sec:appHDE}
In this paper we have revisited the HDE with Ricci scale model presented
by \citet{delCampo13} and \citet{Cardenas:2013moa} but we have taken into account the radiation component.
We consider a flat FLRW universe with DM, HDE, and radiation.
By assuming the possibility of interaction between the dark components,
the dynamics of this universe is governed by the system
\begin{subequations}
\begin{align}
H^{2}&=\frac{8\pi G}{3} \left(\rho_{DM} + \rho_{HDE} + \rho_{r} \right),\\
\dot{\rho_{DM}}&+3H\rho_{DM}=Q,\\
\dot{\rho_{HDE}}&+3H(1+\omega)\rho_{HDE}=-Q,\\
\dot{\rho_{\gamma}}&+4H\rho_{\gamma}=0 \label{eq:rhorad},
\end{align}
\end{subequations}
such that the total energy, $\rho = \rho_{DM} + \rho_{HDE}+ \rho_{\gamma}$, is conserved.
Here, $\omega \equiv p_{HDE}/\rho_{HDE}$ is the EoS parameter of the HDE and
$p_{HDE}$ is the pressure associated with the holographic component.
We may write the holographic energy density as
\begin{equation}
\rho_{HDE}=\frac{3c_{H}^{2}M_{p}^{2}}{L^{2}},
\label{eq:rhoHDE}
\end{equation}
where $L$ represents the IR cutoff scale and $M_{p}$ is the reduced Planck mass.
In the HDE model it is assumed that the energy in a given box should not exceed the
energy of a black hole of the same size. This means that $L^{3} \rho_{HDE}\leq M_{p}^{2} L$,
in this context the numerical constant $c_{H}$ in Eq. (\ref{eq:rhoHDE}) is related with the degree of
saturation of the previous expression. Here, we consider the Ricci scalar, $\mathcal{R}$, as the IR cutoff, i.e.,
$L^{2}=6/\mathcal{R}$, where $\mathcal{R}\equiv 6(2H^{2}+\dot{H})$ \citep{Gao:2009,delCampo13}, then
\begin{equation}
\rho_H= 3\,c_{H}^2\,M_p^2 \,\frac{\mathcal{R}}{6} = \alpha\left(2H^{2} + \dot{H}\right), \label{rhR}
\end{equation}
where $\alpha = 3c_{H}^{2}/8\pi G$.
By defining $x\equiv\nicefrac{\rho_{DM}}{\rho_{HDE}}$ and $y\equiv\nicefrac{\rho_{\gamma}}{\rho_{HDE}}$,
the Friedmann and Raychauduri equations can be rewritten as
\begin{equation}
H^{2}=\frac{c_{H}^{2}}{\alpha} \rho_{HDE}(1+x+y),
\label{eq:HHDE}
\end{equation}
\begin{equation}
\dot{H} = - \frac{3}{2}H^{2}\left(1 + \frac{y}{3(1+x+y)}+\frac{\omega}{1+x+y}\right),
\label{dH}
\end{equation}
The substitution of Eqs. (\ref{eq:HHDE}) and (\ref{dH}) in (\ref{rhR}) leads to the condition
\begin{equation}
\frac{2}{c_{H}^{2}}=1 + x - 3\omega.
\end{equation}
This expression can be evaluated at $a=1$ to obtain $C_{1}=1 + x_{0} -3\omega_{0}$,
where $x_{0}=\nicefrac{\Omega_{DM0}}{\Omega_{HDE0}}$, $\Omega_{HDE0}=1-\Omega_{DM0}-\Omega_{\gamma0}$,
$\Omega_{DM0}$ and $\Omega_{HDE0}$ are the current dark matter and HDE density parameters respectively.
Thus, $x=x_{0}+3(\omega-\omega_{0})$.

On the other hand, the differentiation of $y$ with respect to the cosmological time $t$ yields
\begin{equation}
\dot{y}=\left(\frac{\dot{\rho_{\gamma}}}{\rho_{\gamma}}-\frac{\dot{\rho_{HDE}}}{\rho_{HDE}} \right).
\label{eq:yp}
\end{equation}
We take the time derivative of the Eq. (\ref{rhR}) to obtain
\begin{equation}
\dot{\rho_{HDE}}=\alpha \left( 4H\dot{H}+\ddot{H}\right).
\label{ec:rhohp}
\end{equation}
Similarly, the differentiation of the Eq. (\ref{dH}) results in:
\begin{equation}
\frac{\ddot{H}}{\dot{H}}=-3H \frac{C_{1}+4\omega+\nicefrac{4}{3}\,y}{C_{1}+3\omega+y}
+\frac{4\dot{\omega}+\nicefrac{4}{3}\,\dot{y}}{C_{1}+4\omega+\nicefrac{4}{3}\,y}
-\frac{3\dot{\omega}+\dot{y}}{C_{1}+3\omega+y}.
\label{eq:ddH}
\end{equation}
\noindent
By combining the Eqs. (\ref{eq:rhorad}), (\ref{eq:yp}), (\ref{ec:rhohp}), and (\ref{eq:ddH}),
we obtain the following differential equation for $y$
\begin{equation}
y'=y\frac{3\omega'-C_{1}}{3\omega+C_{1}},
\label{eq:yprime}
\end{equation}
where $'$ stands the derivative with respect to e-foldings $N=\ln a$, i.e., $'=\nicefrac{d}{dN}$.
In the $N$-space, the Eq. (\ref{dH}) reads as
\begin{equation}
H'=-\frac{3}{2}H\left(\frac{C_{1}+4\omega+\nicefrac{4}{3}\,y}{C_{1}+3\omega+y} \right).
\label{eq:Hprime}
\end{equation}
For the EoS CPL parametrization (Eq. \ref{eq:wcpl}), the system of equations \ref{eq:yprime}-\ref{eq:Hprime},
has the following analytical solution
\begin{equation}
E(z)=\frac{1}{\sqrt{1+x_{0}+y_{0}}}\left(1+z\right)^{2}\left(\frac{1+x_{0}}{f}\right)^{\alpha}
\sqrt{f+y_{0}\left(\frac{f}{1+x_{0}}\right)^{2\alpha}},
\end{equation}
where $y_{0}=\nicefrac{\Omega_{DM0}}{\Omega_HDE0}$, $f=1+x_{0}+ z(1+3w_{1}+x_{0})$,
and the exponent $\alpha$ is given by
\begin{equation}
\alpha=1-\frac{3(w_{0}+w_{1})}{2(1+3w_{1}+x_{0})}.
\end{equation}
Thus, we present a new analytical solution of $E(z)$ for the
Ricci HDE model with CPL parametrization including the radiation fluid.\\

\end{document}